\documentclass[usegraphicx,usenatbib,useAMS]{mn2e}

\usepackage{times} 
\usepackage{aas_macros}     
\usepackage{amssymb}        

\newcommand{\bc}{\begin{center}}
\newcommand{\ec}{\end{center}}

\title[Numerical resolution limits on subhalo abundance matching]
{Numerical resolution limits on subhalo abundance matching}
\author[Qi Guo \& Simon White]
       {\parbox{18cm}{Qi Guo$^{1,2}$
           \thanks{Email:qi.guo@durham.ac.uk}, Simon White$^{3}$}
       \\     
       \\
       $^{1}$Partner Group of the Max-Planck-Institut f\"ur
       Astrophysik, National Astronomical Observatories, Chinese
       Academy of Sciences, \\Beijing, 100012, China \\
       $^2$ Institute for Computational Cosmology, Department of
        Physics, University of Durham, South Road, Durham, DH1 3LE, UK
        \\ 
        $^{3}$ Max Planck Institut f\"ur                                        
         Astrophysik, Karl-Schwarzschild-Str. 1, 85741 Garching,
         Germany\\  
      }

\begin{document}
\date{Accepted  ???? ??. 2013 ???? ??}
\pagerange{\pageref{firstpage}--\pageref{lastpage}} 
\pubyear{2013}
\maketitle
\label{firstpage}

\begin{abstract}
 Subhalo abundance matching (SHAM) inserts galaxies into dark matter
 only simulations of the growth of cosmic structure in a way that
 requires minimal assumptions about galaxy formation. A galaxy is
 placed at the potential minimum of each distinct self-bound subhalo
 with a luminosity which is a monotonically increasing function of the
 maximum mass (or circular velocity) attained over the subhalo's
 earlier history. Galaxy and subhalo properties are linked by matching
 model and observed luminosity functions.  Simulated structures can
 then be compared in detail with observation, for example, through
 galaxy correlation statistics, group catalogues, or galaxy-galaxy
 lensing. Robust astrophysical conclusions can be drawn from such
   a comparison only on scales which are unaffected by the numerical
   limitations of the simulation.  Here we compare results for the
 Millennium Simulation (MS) with subhalos defined using the SUBFIND to
 those obtained applying identical analysis to the much higher
 resolution Millennium-II. Correlation statistics on scales between
 200~kpc and 2~Mpc converge to within 20\% only for subhalos with
 masses at infall corresponding to at least 1000 simulation
 particles in the MS. Numerically converged results can be obtained to
 much lower infall particle number, if galaxies are followed even
 after their associated subhalos have been tidally disrupted, as in
 most recent semi-analytic galaxy formation simulations. This allows
 robust comparison between simulation and observation over a wider
 dynamic range in mass than for a SHAM analysis which ignores such
 ``orphan'' galaxies.

\end{abstract}

\section{Introduction}
The $\Lambda$CDM model has proved consistent with observed cosmic
structure over a wide range of scales and epochs, for example, the
fluctuations in the cosmic microwave background \citep{Dunkley2009},
the Lyman $\alpha$ forest absorption in quasar spectra
\citep{McDonald2006} and the large-scale clustering of galaxies
\citep{Percival2010}. In this scenario, small halos form first and
then progressively accrete and merge to form larger systems, a process
which is quite well described by the simple excursion set model of
\cite{Press1974} and its later extensions \citep{Bond1991, Bower1991,
  Lacey1993, Sheth1999}. Current N-body simulations can follow dark
halo growth at high resolution and in its full cosmological context
for objects ranging from halos too small to host galaxies
\citep[e.g.][]{Gao2004} to those hosting the most massive galaxy
clusters \citep[e.g.][]{Gao2012, Angulo2012}. While there is good
agreement on the structure predicted in the absence of baryonic
effects, this agreement evaporates when such effects are included, and
there is little consensus on the properties of the galaxy which should
form in any particular halo \citep[see, for example, the review
  of][]{Frenk2012}. This is because galaxy formation involves complex
and poorly understood processes linking a very wide range of scales,
in particular, star and black hole formation and the strong feedback
processes which they generate.

In the standard scenario, galaxies form as gas cools and condenses at
the centres of the evolving population of dark halos
\citep{White1978}.  The most direct way to study galaxy formation in a
$\Lambda$CDM cosmology is via full simulation
\citep[e.g.][]{Navarro1994, Cen2000, Springel2003, Keres2005,
  Pfrommer2006, Governato2012, Guedes2012}. Such hydrodynamic
simulations follow the gas dynamics in considerable detail, but they
are expensive in terms of both memory and CPU time, and their results
depend strongly on the recipes adopted for unresolved processes
\citep[e.g.][]{Scannapieco2012}. An alternative is provided by
so-called semi-analytic models which employ simple, physically and
observationally motivated prescriptions for baryonic physics
\citep{White1991, Kauffmann1993, Cole1994}. These can be implemented
on halo or subhalo merger trees extracted from dark-matter-only
simulations and are useful for exploring evolution of the galaxy
population \citep{Kauffmann1999, Springel2001, Hatton2003, Kang2005,
  Bower2006, Guo2011}.  Many observables can be predicted rapidly for
large number of galaxies, enabling a detailed check of a model's
ability to reproduce the abundance, clustering and intrinsic property
distributions of galaxies over the full range of observed redshifts,
while simultaneously providing an estimate of the efficiency of each
of the processes involved \citep{Guo2011, Henriques2013}. Such
simulations are particularly useful for calibrating and interpreting
large observational surveys
\citep[e.g.][]{Scoville2006,Coil2007,Meneux2008,Guzzo2008,McCracken2010,Adami2009,Torre2010,Knobel2012,Li2011,Planck2012}. Their
main limitations arise from the simplicity of the individual recipes
adopted, and from the very limited information they provide about the
internal structure of galaxies.

In recent years, subhalo abundance matching (SHAM) has become a
popular method for modeling the large-scale distribution of galaxies
\citep[e.g.][]{Vale2004, Conroy2006, Shankar2006, Baldry2008,
  Moster2010, Guo2010, Behroozi2010, Wake2011}. The main idea is to
place galaxies in dark matter subhalos assuming a monotonic relation
between the stellar mass (or luminosity) of the galaxy, and the
maximum mass (or maximum circular velocity) ever attained by the
subhalo during its earlier history. This is a strong assumption which
is almost certainly incorrect in detail -- for example, the properties
of the galaxy should depend on the redshift at which its subhalo
attains its maximum mass as well as on the mass itself
\citep{Wang2006, Wake2011, Yang2012} -- but the method is convenient
because it requires minimal assumptions and avoids the need to treat
the physics of galaxy formation explicitly. A consequence, of course,
is that it gives relatively little information about this
  physics.  For example, the stellar mass function of galaxies is
reproduced by construction, and so does not directly constrain the
efficiencies of star and black hole formation and of the associated
feedback. Galaxy formation information can, however, be backed out
from the derived stellar mass--halo mass relation and by requiring the
relations obtained at different redshifts to be consistent with the
growth of structure expected in the $\Lambda$CDM model
\citep[e.g.][]{Conroy2009, Moster2013}.  Validating the SHAM approach
requires comparisons with independent observations, e.g., galaxy
correlation functions
\citep[e.g.][]{Conroy2006,Moster2010,Trujillo2011}, group catalogues
\citep[e.g.][]{More2010,Hearin2012}, galaxy-galaxy lensing and
satellite abundances
\citep[e.g.][]{Guo2010,Leauthaud2012,Reddick2013}. Overall the
agreement is quite good, suggesting that the fundamental assumption is
not seriously in error.

Robust simulation-based predictions of galaxy clustering require that
the purely numerical limitations of the simulations be understood.  In
the case of SHAM, the most critical aspect is the need to correctly
identify dark matter subhalos within larger mass objects over the full
period for which they are supposed to represent galaxies within the
corresponding group or cluster.  Subhalos lose mass rapidly once they
merge into a larger system, and any particular simulation may lose
track of a subhalo well before its galaxy should be lost to disruption
or merging.  The extent of such premature subhalo loss depends both on
simulation parameters (primarily mass resolution and softening) and on
the subhalo identification algorithm, and is manifest in a SHAM
analysis through an underprediction of small-scale clustering. It can
be compensated by continuing to follow ``orphan'' galaxies that have
lost their subhalo, as is done in many semi-analytic models, but at
the cost of adopting parametrised recipes for the lifetimes and orbits
of the galaxies, thus losing some of the simplicity of the SHAM
methodology \citep{Moster2010, Neistein2011, Moster2013, Neistein2012}.

Here we study these resolution effects by comparing results from SHAM
analyses applied to the Millennium \citep[MS,][]{Springel2005} and
Millennium-II \citep[MS-II,][]{Boylan2009} simulations. These differ
by a factor of 125 in particle mass and a factor of 5 in force
softening, allowing us to explore how well the clustering predicted
for the MS converges to that predicted for the much higher resolution
MS-II. In addition, we illustrate how this convergence is affected by
including orphan galaxies by applying a directly analogous
clustering analysis to the semi-analytic galaxy catalogues made for
the two simulations by \cite{Guo2011}. Note that we are not here
concerned with how well either scheme reproduces the clustering of observed
galaxies (indeed, \cite{Guo2011} show explicitly that their model
fails to match observation in several aspects) but rather with
understanding the mass range over which each scheme produces
{\it numerically converged} results when applied to the MS.  This is clearly
a prerequisite for robust conclusions about astrophysics to be drawn
from a comparison with real galaxy clustering.


In contrast, \cite{Wetzel2010} discuss the requirements on subhalo
tracking for the SHAM approach to reproduce the {\it observed}
clustering of SDSS galaxies with M$_r<$-20.5, concluding that the
relevant subhalos must be followed at least to the point where their
mass has dropped by a factor of 30 to 100. They note that since
typical simulations are able to identify subhalos down to a limit of
about 30 particles, this implies that the halos identified with 
SDSS galaxies must be resolved with 1000 to 3000 particles at infall
if the observed clustering is to be matched. This is similar to the
criterion which we derive below for convergence of subhalo clustering
to the results of a much higher resolution simulation, but
since \cite{Wetzel2010} do not consider convergence issues, and we do
not consider matching to observation, these two limits are, in fact,
conceptually quite distinct. Some discussion of numerical issues was
already given earlier by \cite{Wetzel2009} although without detailed
convergence tests.

We summarize the main features of the MS and the MS-II in Sec.
~\ref{sec:sim}, while in Sec.~\ref{sec:sam} we briefly describe the
semi-analytic model which \cite{Guo2011} implemented on the two
simulations. In Sec.\ref{sec:abundance} we study how the abundance of
subhalos as a function of mass at infall and of maximum circular
velocity at infall is affected by the limited mass resolution of the
MS, while in Sec.~\ref{sec:cor} we examine how this affects the
small-scale clustering of subhalos of given abundance, finding
substantial effects for infall masses corresponding to fewer than
about 1000 particles. Sec.~\ref{sec:samresults} then uses
the semi-analytic model to illustrate how the inclusion of orphan
galaxies can substantially increase the mass range over which
numerically converged clustering results can be obtained.
Sec.~\ref{sec:con} recapitulates and discusses our main results.

\section{Simulations}
In this section, we summarize the main features of the two simulations
used in our analysis, and we briefly describe the semi-analytic models
used to generate the simulated galaxy catalogues we compare with our
SHAM catalogues.
\subsection{The MS and MS-II}
\label{sec:sim}

This analysis is based on two cosmological simulations, the Millennium
and the Millennium-II. Both assume a $\Lambda$CDM cosmology with
parameters based on a combination of data from the 2dFGRS
\citep{Colless2001} and the first-year WMAP release
\citep{Spergel2003}: $\Omega_m$ = 0.25, $\Omega_{\Lambda}$ = 0.75,
$\Omega_b$ = 0.045, n = 1, $\sigma_8$ = 0.9 and h = 0.73. These
parameters are at best marginally consistent with the nine-year WMAP
results\citep{Hinshaw2012}. However, this is of no consequence for the
analysis we present in this paper.

The MS and the MS-II were carried in periodic volumes of side 500$\rm
h^{-1}$Mpc and 100$h^{-1}$Mpc, respectively. Both simulations used
2160$^3$ particles to follow the dark matter distribution from
redshift 127 to the present day. Comoving gravitational softening
  lengths of 5kpc and 1kpc were used for the MS and the MS-II,
  respectively, so that the ratio of softening length to the mean
  interparticle particle separation was constant and was identical in
  the two simulations. Both the softenings and the integration
  parameters were chosen sufficiently conservatively that structural
  convergence between the two simulations is good and differing mass
  resolution is the primary driver of the effects we investigate below
  \citep[see, for example, the tests presented by][]{Boylan2009} 64
snapshots were stored for the MS and 68 for the MS-II, with the last
60 being identical in the two simulations. At each output time,
friends-of-friends (FOF) groups were identified by linking particles
with separation less than 0.2 of the mean value \citep{Davis1985}. The
SUBFIND algorithm \citep{Springel2001} was then applied to identify
all self-bound subhalos within each FOF halo, and every subhalo was
linked to a unique descendent at the next later output time in order
to construct merger trees describing the complete assembly history of
each $z=0$ subhalo.  Halo/subhalo data were stored only for objects
containing at least 20 particles, so that the smallest resolved
subhalo in the MS has mass 2.4$\times 10^{10}M_{\odot}$, while the
corresponding resolution limit in the MS-II is 1.9$\times
10^8M_{\odot}$. The readers are referred to \cite{Springel2005}and
\cite{Boylan2009} for more detailed descriptions of the two
simulations.

In each FOF group, the most massive self-bound subhalo is referred to as
the ''main subhalo'' or ''halo''. The center of the main subhalo is taken to be
the gravitational potential minimum of the FOF group.  The virial radius $R_{vir}$ of
the main subhalo is defined as the maximum radius within which the
mean density is 200 times the critical value. The total mass enclosed
is then defined as the virial mass of the group, $M_{vir}$.  Within the
virial radius, the maximum circular velocity is defined as
\begin{equation}
V_{\rm max} = {\rm max}_r\big\{\sqrt{GM(r)/r}\big\}.
\end{equation}
$V_{\rm max}$ is a useful indicator of potential well depth for subhalos,
since it is less affected by stripping and boundary definition effects
than subhalo mass.

In the standard galaxy formation scenario, galaxy properties are
expected to be closely related to the mass and potential well depth of
the halo in which they form.  When a halo falls into a more massive
system, it progressively loses its outer regions until it is
eventually destroyed.  The galaxies, however, are much more compact,
and so are little affected by tides until the final stage of
disruption/merging.  The baryonic properties of satellite galaxies are
thus expected to be more closely related to the properties of their
subhalo at the time of infall than to its current properties. In this
paper, we follow the standard procedure in recent SHAM studies, and
characterise subhalos by their mass or maximum circular velocity at
infall (defined as the latest time when they were the main subhalo of
their FOF group) when estimating their abundance for matching
purposes. For the main subhalo of each group, this is the current value
of these quantities. In both cases, however, we refer to the
quantities as $M_{\rm max}$ and $V_{\rm max}$, respectively, even though the
past maximum mass of a satellite subhalo does not always occur at the
time of infall.

\subsection{Semi-analytic Models}
\label{sec:sam}

The semi-analytic simulation technique populates halos with galaxies
by applying a set of physically or observationally motivated recipes
to merger trees extracted from N-body simulations. In this paper we
will use galaxy catalogues from the semi-analytic model of
\cite{Guo2011} which was implemented simultaneously on the same MS and
MS-II subhalo catalogues which we use for our SHAM modelling. This
semi-analytic model follows the reionization, infall, shock-heating,
radiative cooling and condensation of diffuse gas onto galactic disks;
the star formation within these disks and the associated metal
production and wind generation; disk instabilities and galaxy mergers
with their associated starbursts, formation of bulges, and formation
and fuelling of central black holes; and the feedback from these black
holes into the surrounding hot gas atmospheres. The model successfully
reproduces the properties, abundances and large-scale clustering of present-day
galaxies ranging from dwarf spheroidals to giant cDs. In this model
all subhalos have a galaxy at their centre but not all galaxies are
associated with a subhalo. When a subhalo is tidally disrupted, a
timescale for orbital decay is calculated from the position of the
subhalo just prior to disruption and from the masses of the subhalo,
of its central galaxy and of the main halo within which it orbits. The
position of the associated ``orphan'' galaxy relative to main halo
centre is then taken to be that of the most bound particle of the
pre-disruption halo, shrunk according to a simple dynamical friction
model so that a merger occurs one dynamical friction time after
subhalo disruption. Details may be found in \cite{Guo2011}.

\section{Results}
\subsection{Halo/subhalo abundance}
\label{sec:abundance}

\cite{Boylan2009} tested for numerical converence between the MS and
the MS-II by comparing a number of standard quantitative measures of
structure in the mass distributions of the two simulations. In
particular, they calculated and compared mass autocorrelation
functions at redshifts $z=0, 1, 2$ and $6$ finding agreement to better
than about 10\% on physical scales between $20h^{-1}$kpc and
$1.0h^{-1}$Mpc and to better than 20\% out to a comoving scale of $10
h^{-1}$Mpc. They also compared FOF halo mass functions at $z = 0$ and
$z = 6$, again finding agreement to within about 10\% above a mass of
4 $\times 10^{10}M_{\odot}$ (around the minimum halo mass resolved in
the MS). A similar test was carried out by \cite{Angulo2012} who in
addition compared results with the much larger Millennium-XXL
simulation.

\begin{figure} \bc \hspace{-0.6cm}
\resizebox{8cm}{!}{\includegraphics{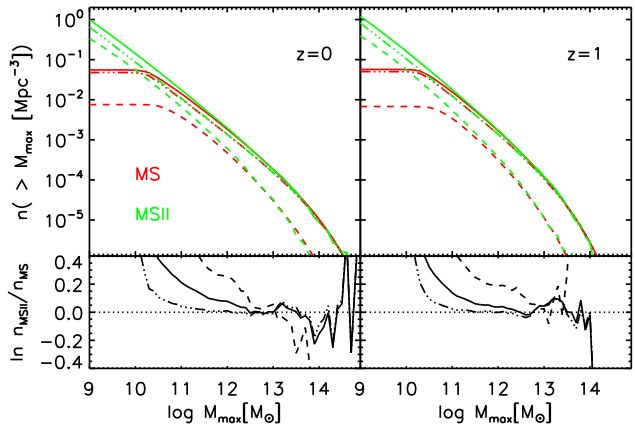}}\\%
\caption{Upper panels: Cumulative abundance of subhalos as a function
  of their mass at infall. Red and green curves are results from the
  MS and the MSII respectively. Lower panels: ratios of the MS and
  MS-II abundances shown in the upper panels. Left panels are for $z =
  0$ and right for $z = 1$.  In all panels, solid curves denote the
  full subhalo abundance, while dashed and dot-dashed curves separate
  out the contributions from satellite subhalos and main subhalos
  respectively.}
\label{fig:MF} \ec

\end{figure}

Both these statistics depend primarily on the abundance, internal
structure and spatial distribution of halos and are relatively
insensitive to substructure. This explains why good results are
obtained down to masses corresponding to only a few tens of
particles. Much larger resolution effects are found for statistics
which depend on the properties of subhalos.  Fig.~\ref{fig:MF}
illustrates this point for the particular case of the subhalo
abundance as a function of infall mass, a statistic often used
in subhalo abundance matching (SHAM) analyses. The overall abundances
are shown by solid curves, with the contributions from main subhalos
and satellite subhalos indicated by dot-dashed and dashed curves
respectively. The main subhalo mass functions agree to within 10\%
between the two simulations for masses above a few $10^{10}M_{\odot}$,
just as for FoF halo masses, but differences of this order appear for
satellite subhalos already at masses one hundred times greater. As a
result, the overall subhalo mass function of the MS agrees with that
of the MS-II to within 10\% only  above $10^{12}M_{\odot}$ at
$z=0$, and above $2\times10^{11}M_{\odot}$ at $z=1$. Thus, SHAM assignment of
galaxies to simulations will only give numerically converged results
at the 10\% level for infall masses corresponding to $\sim 10^3$
particles or more. 

\begin{figure} \bc \hspace{-0.6cm}
\resizebox{8cm}{!}{\includegraphics{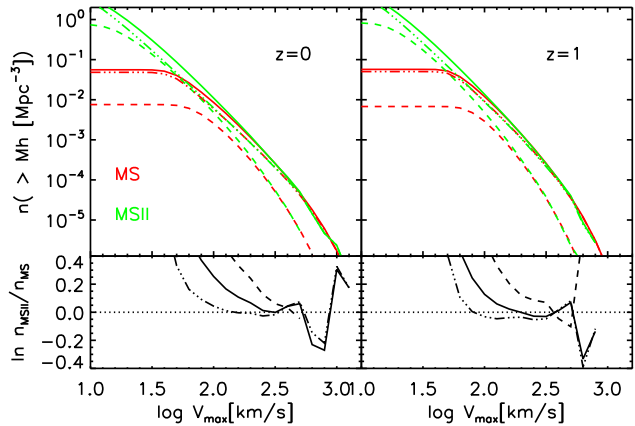}}\\%
\caption{Similar to Fig. 1 but for subhalo abundances as a function of
  the value of peak circular velocity at infall, $V_{\rm
    max}$.}  \label{fig:Vmax} \ec
\end{figure}

SHAM analyses often consider subhalo abundance as a function of
$V_{\rm max}$ rather than $M_{\rm max}$, since $V_{\rm max}$ is a
useful indicator of subhalo potential well depth and is less sensitive
to the details of subhalo identification than $M_{\rm max}$.
Fig.~\ref{fig:Vmax} is identical in format to Fig.~\ref{fig:MF} except
that abundances are shown as functions of $V_{\rm max}$. The change in
parametrisation has little effect on the convergence behaviour. The
abundances of main subhalos agree within 10\% for $V_{\rm max}$ values
above about 80~km/s, but similar agreement is obtained for satellite
subhalos only for $V_{\rm max}> 300$~km/s.  As a result a SHAM
analysis based on the MS is converged at the 10\% level only for
galaxies with rotation velocities similar to the Milky Way.
 
\subsection{Subhalo correlations}
\label{sec:cor}
\begin{figure*}
\bc
\hspace{-0.6cm}
\resizebox{17cm}{!}{\includegraphics{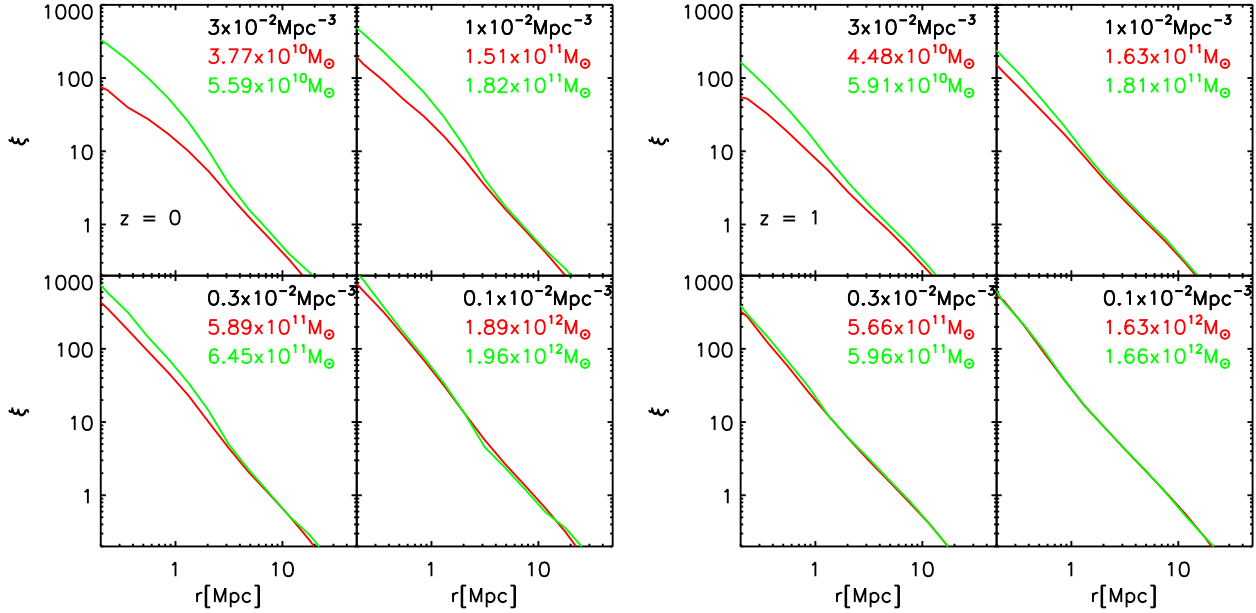}}\\%
\caption{Two-point autocorrelation functions for subhalo samples
  defined above various lower limits on $M_{\rm max}$, the subhalo
  mass at infall.  As in Fig. 1, red curves are for MS samples, green
  curves for MS-II samples. The left set of panels shows data for
  $z=0$ while the right set are for $z=1$. Each panel compares results
  for samples with the abundance shown in black in its upper right
  corner. The corresponding lower limits in $M_{\rm max}$ are also
  shown, with colour indicating the relevant simulation. Note that
  these mass limits do not agree precisely beause of the lack of
  convergence seen in Fig.~\ref{fig:MF}.}
\label{fig:cormassz0}
\ec

\end{figure*}

As the last subsection shows, differences in subhalo abundance between
the two Millennium Simulations are primarily a reflection of lack of
convergence in the properties of low-mass satellite subhalos. As a
result, one may expect the small-scale clustering of subhalos to show
even more serious disagreement.  This is explored in
Fig.~\ref{fig:cormassz0} which compares autocorrelation functions for
mass-limited subhalo samples chosen to have exactly matched abundances
in the two simulations. The four panels in each set correspond to
abundances differing by factors of three, as noted by the black label
in each panel. The red and green labels indicate the lower limits on
infall mass for the MS and MS-II respectively, with the same
colours used for the corresponding curves.  The left set of panels
shows results for $z = 0$ while the right set shows results for
$z=1$. We use this particular matching scheme in order to indicate the
expected disagreement in galaxy correlations if the two simulations are
populated with galaxies using SHAM.

The highest subhalo abundance considered ($n=3\times 10^{-2}{\rm
  Mpc}^{-3}$) corresponds to a lower limit on infall halo mass of
about $5\times 10^{10}{\rm M}_{\odot}$ both at $z=0$ and at
$z=1$. This is about the mass of the halo thought to host the Small
Magellanic Cloud \citep[e.g.][]{Guo2010}.  To this limit the
clustering amplitude of subhalos at $z=0$ is lower in the MS than in
the MS-II by 20\% on large scales, by a factor of two at $r_p=2$~Mpc,
and by a factor of about four on 100~kpc scales. The discrepancy is
similar on large scales at $z=1$, but only about half as big on Mpc
scales and below. For sparser samples, corresponding to higher minimum
values of $M_{\rm max}$, the agreement improves, but only for the
sparsest sample ($n=10^{-3}{\rm Mpc}^{-3}$, $M_{\rm max}>2 \times
10^{12}{\rm M}_{\odot}$) is it better than 20\% on all but the
smallest scales.  Thus at the resolution of the MS, the SHAM procedure
gives ``precision'' results for present-day clustering only for
subhalos for which $M_{\rm max}$ is greater than about $10^3$ times
the mass of a simulation particle, corresponding to galaxies more
massive than the Milky Way. A similar requirement for numerical
convergence was suggested by \cite{Wetzel2009} although without
presentation of the supporting numerical results. 

At $z=1$ the convergence is already quite good for $M_{\rm max}$
values corresponding to about 300 particles. This difference with
$z=0$ would not be present in a universe where clustering was
self-similar (i.e. an Einstein-de Sitter universe with a power-law
initial fluctuation spectrum) and where the simulations compared at
different redshifts have similar values of $M_*(z)/m_p$, where
$M_*(z)$ is the characteristic nonlinear mass of clustering at
redshift $z$ and $m_p$ is the simulation particle mass. As a result,
the difference we find must reflect one of three things: (i) the lower
value of $M_*(z)/m_p$ at $z=1$; (ii) the fact that the dimensionless
growth rate of structure is lower at $z=0$ than at $z=1$; or (iii) the
fact that the effective index of the linear power spectrum $M_*(z)$ is
larger at $z=0$ than at$z=1$. Differentiating between these three
possibilities would require additional simulation material.

\begin{figure*}
\bc
\hspace{-0.6cm}
\resizebox{17cm}{!}{\includegraphics{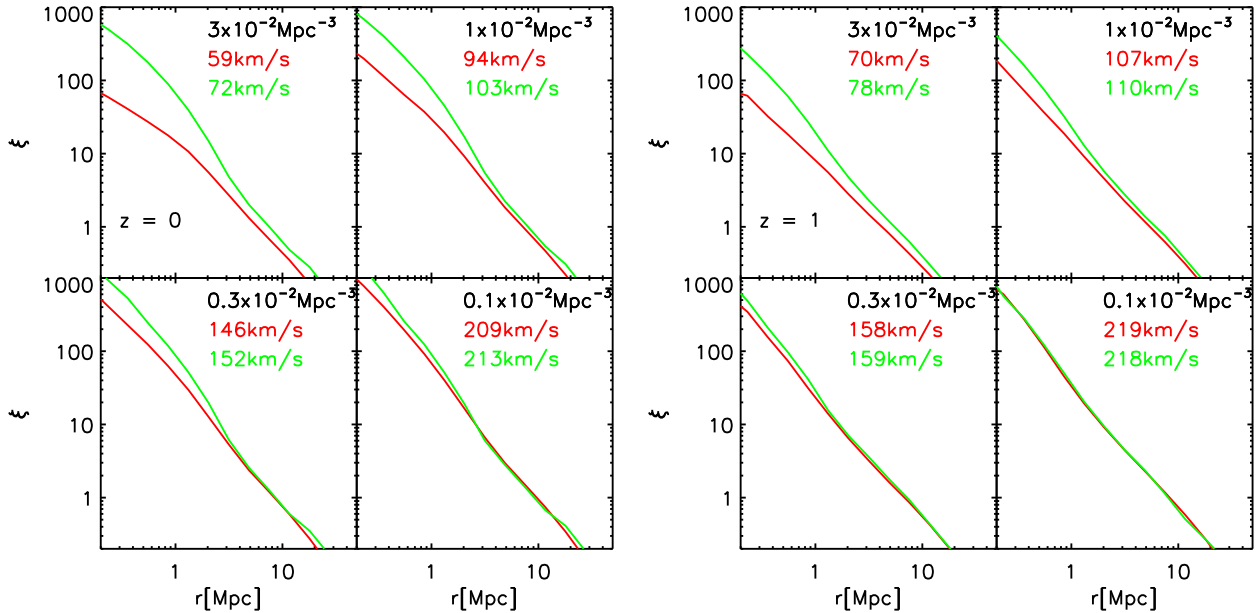}}\\%
\caption{Similar to Fig.~\ref{fig:cormassz0} but here for samples
  selected above a threshold in $V_{\rm max}$, the value of
  maximum circular velocity at subhalo infall.
  Coloured labels in each panel now show the values of $V_{\rm max}$ at
  the thresholds which correspond to the chosen abundance in each of the
  two simulations. They do not agree because of the lack of
  convergence seen in Fig.~~\ref{fig:Vmax}}
\label{fig:corvmaxz0}
\ec

\end{figure*}

As shown in Fig.~\ref{fig:corvmaxz0}, very similar discrepancies are
seen if subhalo samples are defined above a limiting value of $V_{\rm
  max}$ rather than $M_{\rm max}$. Indeed, comparison with
Fig.~\ref{fig:cormassz0} shows the agreement to be significantly
worse, both at $z=0$ and at $z=1$, when this supposedly better proxy
for galaxy mass is used for abundance matching in the two
simulations. At $z=0$ the ratio reaches a factor of eight at a
separation of 200 kpc for the highest density samples. The large-scale
discrepancy is also bigger than in samples selected by $M_{\rm max}$. This
is because at given mass (sub)halos formed at higher redshift are more
compact and hence have higher $V_{\rm max}$. Thus, given that satellites
typically assembled earlier than centrals of similar $V_{\rm max}$ or
$M_{\rm max}$, ranking by the former rather than the latter increases the
ratio of satellite to central subhalos at each abundance.

A comparison of Figs~\ref{fig:cormassz0} and \ref{fig:corvmaxz0} shows
that, as already discussed by \cite{Reddick2013}, SHAM clustering
predictions depend significantly on which matching scheme is adopted.
At the smallest scale plotted, the predictions for a given simulation
and abundance can differ by almost a factor of two. Clearly at most
one of the two schemes could be a precise representation of the
small-scale clustering of any particular observational sample.

\subsection{Autocorrelations for SA galaxies}
\label{sec:samresults}
Our semi-analytic model follows satellite galaxies from the time their
associated subhalos are disrupted until the galaxies themselves are
tidally destroyed, or they merge with a more massive system. If the
orbits of such orphan galaxies are modelled adequately, the galaxy
population should be less subject to convergence problems of the kind
discussed above than the subhalo population we have focussed on so
far. \cite{Guo2011} show that galaxy abundances agree to better than
10\% between the MS and the MS-II down to stellar masses of about
$2.5\times10^{9}{\rm M}_{\odot}$, corresponding to $n\sim 1.2\times 10^{-2}{\rm
  Mpc}^{-3}$. In addition, at given stellar mass, they show galaxy
autocorrelations to agree to similar accuracy over the full separation
range between 100~kpc to 10~Mpc. While clearly encouraging, these
convergence tests are not directly comparable to those we presented
above.  We have therefore created matched, stellar-mass-limited
samples of galaxies from the MS and MS-II both at $z=0$ and at $z=1$
and we compare their clustering exactly as we did for
abundance-matched subhalo samples in Figs.~\ref{fig:cormassz0}
and~\ref{fig:corvmaxz0}.

\begin{figure*}
\bc
\hspace{-0.6cm}
\resizebox{17cm}{!}{\includegraphics{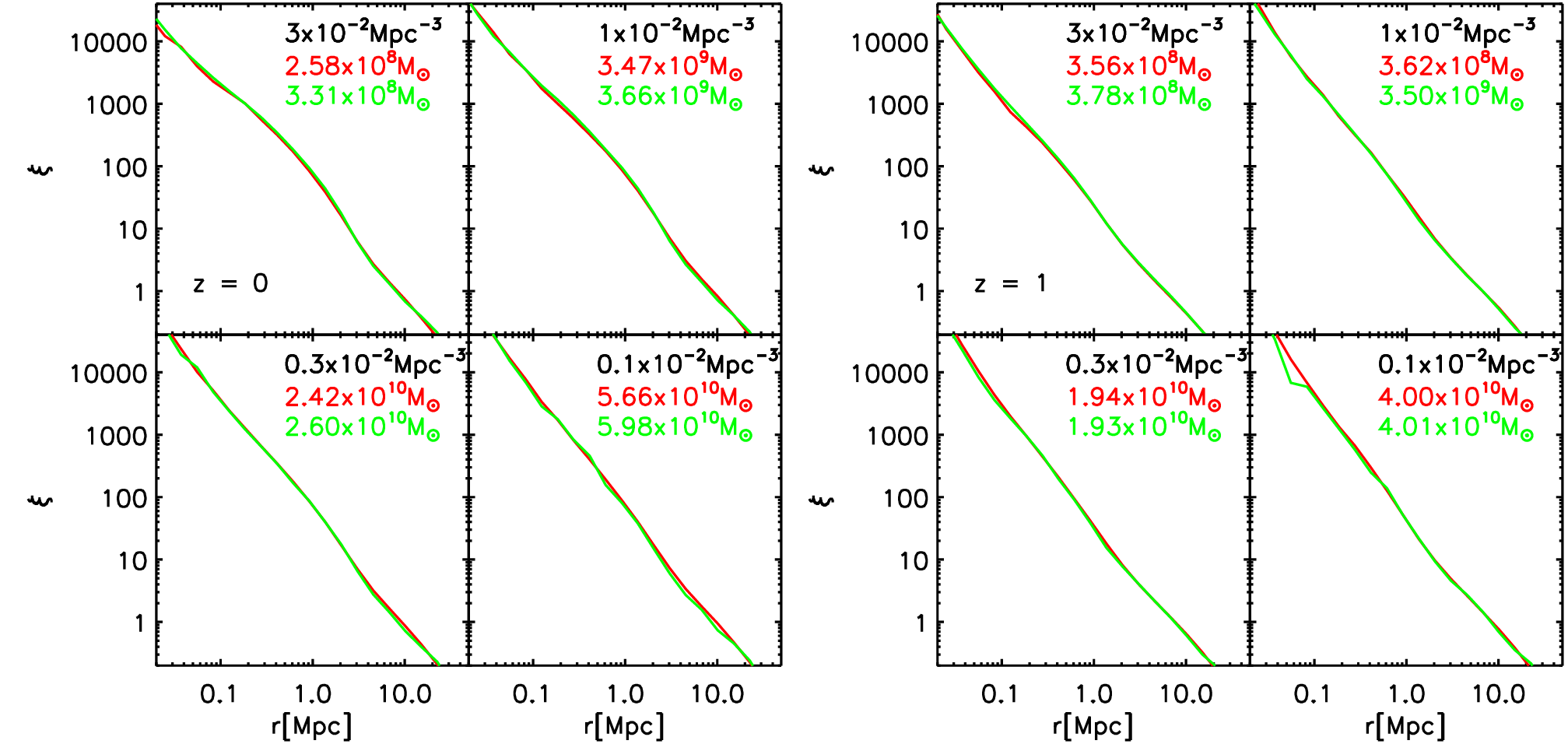}}\\%
\caption{Similar to Fig.~\ref{fig:cormassz0} but for semi-analytic
  galaxy samples selected above thresholds in stellar mass. Coloured
  labels in each panel now show the thresholds required to match the
  chosen abundance in each of the two simulations. They agree quite
  closely because the stellar mass functions of the two simulations
  almost coincide at the relevant masses. In contrast to
  Figs~\ref{fig:cormassz0} and~\ref{fig:corvmaxz0}, the
  autocorrelation functions here agree very well between the two
  simulations.}
\label{fig:cormstarz0}'
\ec
\end{figure*}

The results are shown in Fig. ~\ref{fig:cormstarz0}. A first point to
note is that, except for the densest samples at $z=0$, the stellar
mass limits for the two samples in each panel never differ by more
than about 5\%. This again emphasises how well the stellar mass
functions agree in the two simulations, much better than the
corresponding subhalo abundance functions $n(M_{\rm max})$ and
$n(V_{\rm max})$ (e.g. compare with the differences in threshold
listed in the panels of Figs~\ref{fig:cormassz0}
and~\ref{fig:corvmaxz0}). The most significant point, however, is
that, apart from some noise effects on small scales, the
autocorrelation functions in the two simulations agree almost
perfectly in all panels across the full range of scales plotted. This
is a strikingly different situation from those shown in
Figs~\ref{fig:cormassz0} and~\ref{fig:corvmaxz0} for the
abundance-matched subhalo samples.  The relatively simple
orphan-tracking methods used in the SA models substantially extend the
range in stellar mass for which numerically converged results for
galaxy clustering can be obtained, allowing results from the MS to be
compared robustly with observation over more than two orders of
magnitude in stellar mass.  \cite{Guo2011} found substantial
discrepancies for dwarf galaxies which more recent work has shown to
be partly due to cosmology and partly to galaxy formation physics
\citep{Guo2013, Henriques2013}.

\section{Discussion}
\label{sec:con}

Subhalo abundance matching (SHAM) aims to provide a direct, physical
and relatively assumption-free scheme for using large cosmological
N-body simulations to interpret observational surveys of galaxy
clustering and large-scale structure. Its principal assumption is that
an observable property of the galaxies (for example their $r$-band
luminosity or their stellar mass) can, to sufficient accuracy, be
assumed to be a monotonic function of some dark halo property
(for example the mass or peak circular velocity at infall)
which can be reliably measured for subhalos identified in an N-body
simulation. Abundance matching then allows subhalos in the simulation
to be populated with galaxies with known values of the observable.

Regardless of whether this procedure is an adequate representation of
the galaxy formation process, it will only predict {\it numerically}
converged properties for the galaxy population on scales where the
abundance of subhalos as a function of the chosen property is
independent of parameters such as simulation size and resolution. To
investigate this issue, we compared subhalo catalogues constructed
using the SUBFIND algorithm from the MS and MS-II, two simulations
differing in mass resolution by a factor of 125. We found subhalo
abundances to agree to better than 10\% only for subhalos with infall
masses corresponding to at least $10^3$ particles in the MS.
Discrepancies arise at lower mass because satellite subhalos are
tidally disrupted too early in the lower resolution simulation. This
has a particularly marked effect on the small-scale clustering of
subhalos, and SHAM predictions for galaxy autocorrelation functions at
separations of 1.0~Mpc or less can be in error by factors of two or
more if they use subhalos significantly less massive than the thousand
particle limit.

Using the infall value of peak circular velocity ($V_{\rm max}$) 
rather than of mass ($M_{\rm max}$) to characterize
subhaloes in the SHAM procedure makes numerical
convergence problems worse rather than better. This is because at
given abundance the fraction of satellites is larger in $V_{\rm
  max}$-limited subhalo samples than in $M_{\rm max}$-limited
samples. At the resolution of the MS, the clustering predicted by
either SHAM scheme is not well converged even for galaxies of similar
mass to the Milky Way. Specifically, predicting autocorrelation
functions accurate to 20\% or better requires $M_{\rm max}>
10^{12}M_\odot$ or $V_{\rm max}>200$~km/s.  Note that this conclusion
may depend on the specific algorithm used to identify subhalos. The
study of \cite{Onions2012} suggests that relatively small differences
are likely for the algorithms in frequent general use, and that
SUBFIND is fairly typical. Our own study makes clear that SHAM should
not be used for ``precision'' interpretation of clustering
observations without careful convergence testing of the kind we
present.

Although some early SHAM studies populated subhalos with galaxies down
to $M_{\rm max}$ or $V_{\rm max}$ values where we would predict
substantial resolution effects \citep[e.g][]{Tasitsiomi2004,
  Conroy2006}, most recent work has been quite conservative and does
not appear to violate the limits we suggest here by large
factors.\footnote{An exception is \cite{Behroozi2013a, Behroozi2013b}
  who present results down to subhalo infall masses coresponding to
  fewer than 100 particles, where we expect resolution effects to be
  large.}  Nevertheless, without an explicit test it is difficult to
be sure that results are converged, and comparison of our Fig.~4 to
Figs~14 and 15 of \cite{Trujillo2011} suggests that resolution
may significantly affect their SHAM analysis of the {\it Bolshoi}
simulation for $M_r>-20$. If this is indeed the case, then significant
though relatively small effects are also present in other recent SHAM
analyses of this same simulation at the corresponding abundances and separations
\citep[e.g.][]{Reddick2013, Behroozi2013a, Behroozi2013b, Watson2013}.

Our study of clustering for the semi-analytic galaxy populations of
\cite{Guo2011} shows that tracking orphan galaxies explicitly can
substantially improve numerical convergence at low stellar
mass. Apparently, the treatment of orphans in these models does a
surprisingly good job of removing the resolution dependence of
satellite disruption, the principal cause of discrepancies between the
SHAM catalogues built from the MS and MS-II.\footnote{Note that we are
  claiming only that semi-analytic models are numerically converged to
  significantly lower mass than SHAM models, when both are applied to
  a given simulation. We are {\it not} claiming they are a better
  representation of the real world. Indeed, \cite{Guo2011}
  demonstrated that their model significantly overpredicts the
  clustering of low-mass galaxies.} This has already been recognised
in several previous studies which included semi-analytic modelling of
orphans in an otherwise near-standard SHAM analysis \citep{Moster2010,
  Neistein2011, Neistein2012, Moster2013}.  In practice, however, SA
models have an advantage over SHAM schemes when interpreting
clustering in large observational surveys not just because their
treatment of orphans substantially expands the range of subhalo masses
which can be robustly populated with galaxies, but also because their
physically realistic treatment of galaxy evolution guarantees
consistent assignment of galaxies to subhalos over the relatively
broad range of redshifts spanned by most next-generation surveys.

\section*{Acknowledgments}
GQ acknowledges a Royal Society Newton International Fellowship the
Partner Group program of the Max Planck Society, the National basic
research program of China (program 973 under grant No. 2009CB24901),
and the Young Researcher Grant of National Astronomical Observatories,
CAS, the NSFC grants program (No. 11143005).  SW acknowledges support
from Advanced Grant 246797 “GALFORMOD” from the European Research
Council.

\bibliographystyle{mn2e}

\setlength{\bibhang}{2.0em}
\setlength\labelwidth{0.0em}

\bibliography{draft_v3}

\end{document}